\documentstyle[aps,preprint]{revtex}

\begin{document}

\newcommand{\bena}{\begin{eqnarray}}
\newcommand{\eena}{\end{eqnarray}}
\newcommand{\epsn}{\epsilon}
\newcommand{\taubar}{\bar{\tau}}
\newcommand{\betabar}{\bar{\beta}}
\newcommand{\ubar}{\bar{u}}
\newcommand{\ubarx}{\bar{u}_x}
\newcommand{\ubary}{\bar{u}_y}
\newcommand{\ubarxt}{\bar{u}_{x}(\bar{\tau})}
\newcommand{\ubaryt}{\bar{u}_{y}(\bar{\tau})}
\newcommand{\ombarn}{\bar{\omega}_n}
\newcommand{\Lam}{\Lambda}
\newcommand{\ubarbf}{\bar{\bf u}}
\newcommand{\omz}{\omega_0}
\newcommand{\IHD}{I_{\rm HD}}
\newcommand{\suminf}{\sum^{\infty}_{n=-\infty}}
\newcommand{\qtqz}{Q(T)/Q_0}
\newcommand{\qzqz}{Q(0)/Q_0}
\newcommand{\omztau}{\omega_0 \tau_r}

\draft

\title{
Thermally Assisted Quantum
Vortex Tunneling in the Hall and Dissipative Regime}

\author{
Gwang-Hee Kim$^{1*}$ and Mincheol Shin$^{2}$
}

\address{
${}^{1}$Department of Physics, Sejong University,
Seoul 143-747, Republic of Korea\\
${}^{2}$ School of Engineering, Information and Communications
University, Daejeon 305-714, Republic of Korea
}
\date{Received \hspace*{10pc}}
\maketitle

\thispagestyle{empty}

\begin{abstract}
Quantum  vortex tunneling  is studied for the case where  the
Hall and the dissipative dynamics are simultaneously present. For
a  given  temperature, the magnetization relaxation rate is
calculated as a function of the external current and the
quasiparticle scattering time.
The relaxation rate is solved analytically at zero temperature
and obtained numerically at finite temperatures by the variational
method. In the moderately clean samples, we have found that
a minimum in the relaxation rate exists at zero temperature, which
tends to disappear with increase in the temperature.
\\
\end{abstract}
\pacs{PACS numbers: 74.60.Ge}

\section{introduction}
The depinning properties of vortices in high
temperature superconductors(HTSC) have generated a good deal of
interest over the past decade.\cite{bla94,yes96}
Yeshurun and Malozemoff\cite{yes88} reported
on the existence of the giant flux creep
which arises from the thermally activated motion of vortices from
one metastable state to a neighboring one.
The probability for such a process is proportional to
$\exp (-U /k_B T)$, where $U$ is the height of the energy barrier
which depends on the pinning strength and
the external current.\cite{tin}
At an extremely low temperature the exponent diverges and
the vortex cannot move out of the pinning sites any more.
Hence, the dynamical magnetization relaxation rate $Q$ defined
as $k_B T/U$ is expected to vanish at $T=0$.
However, many experiments\cite{mot,gri,tej,dal,hoe,sef}
have demonstrated that the relaxation rate
does not disappear at sufficiently  low temperatures,
which leads to the existence
of quantum tunneling of vortices trapped in the pinning potential.

In general, quantum vortex  creep is well described by the
dynamics of two major forces: the  Hall force and dissipative force. 
Within the collective pinning theory, Blatter {\it et
al.}\cite{bla91} considered the quantum vortex tunneling for the
case where the dissipative term is dominant in the motion of
vortices. On the other hand, Feigel'man {\it et al.}\cite{fei}
proposed that the Hall tunneling is dominant in clean
superconductors by estimating the low-lying level spacing in the
vortex core and the transport relaxation time of the charge
carriers. Many experimental results have been interpreted within
the two frameworks. Recently, however, van Dalen {\it et al.}\cite{dal}
observed experimentally that  the vortex tunneling in HTSC  may
occur in the intermediate regime between the purely dissipative tunneling and
the superclean Hall tunneling. Feigel'man {\it et
al.}\cite{fei} and Morais Smith {\it et al.}\cite{mor} studied
the problem  in the two  regimes, but they only obtained the
qualitative results based on the scaling analysis of the action.
The main difficulty of the problem is in the fact that there is the time
nonlocality caused by the dissipative dynamics. Recently, the
present authors\cite{kim98}  have treated the problem
quantitatively by using the variational method and presented the
numerical results for the magnetic relaxation rate at zero
temperature in the intermediate  regime. Later, Melikidze\cite{mel}
studied a similar problem by considering the quadratic
Hamiltonian of the vortex coupled to the environment. Through
the analytic diagonalization, he obtained the dynamical magnetization
relaxation rate at zero temperature as a function of the Hall and
dissipative coefficients and found the  minimum  feature  in  the
intermediate regime. Previous works have treated the problem
only at zero temperature, but we extend it to finite temperature in
this work.
Based on  the instanton approach,
we have obtained the numerical results for the relaxation rate at finite temperatures and
its analytic expression around the crossover temperature between thermal
activation and quantum tunneling.
Using the functional
dependence of the relaxation rate on the Hall and dissipative coefficients at the
crossover temperature, we have also obtained the analytic expression of the 
relaxation rate at zero temperature.

This paper is organized as follows. In Sec. II, we introduce
the general formulation for the vortex tunneling rate in the 
presence of the Hall and the dissipative dynamics based
on the instanton method, and discuss the Ohmic dissipation
formulated by Caldeira and Leggett\cite{cal}. In Sec. III, 
we calculate the magnetic relaxation rate by 
taking into account the pinning potential barrier generated by impurities. 
Writing the action
and the corresponding classical equations  in the Fourier space, 
we analytically calculate the relaxation rate 
as a function of the external current 
and the Hall and the dissipative coefficients at zero temperature. We also discuss
the  minimum of the relaxation rate in the intermediate regime.
In Sec. IV, 
we  numerically calculate the finite-temperature relaxation rate
based on the variational method. It is found that the minimum in the
relaxation rate tends to disappear with the increase in the temperature.
We conclude in Sec. V.

\section{Basic formulation}

We consider the pancake
vortex in the $xy$ plane with length $L_c$ along the $z$ axis.
$L_c$ is the collective pinning length which can be expressed in terms
of the mass anisotropy parameter $\varepsilon^{2}_{a}=m/M < 1$, the
coherence length $\xi$, the depairing curremt density  $j_0$, and
the critical current density $j_c$ :
$L_c \simeq \varepsilon_a 
\xi (j_0 /j_c)^{1/2}$, within the weak collective pinning
theory.\cite{bla94} $L_c$ is obtained by minimizing the energy
density which includes the elastic energy of the vortex string,
the energy gain from the random pinning potential and the
contribution from the Lorentz force. Thus, each segment of the length
$L_c$ of the vortex is pinned by the collective action of all
the defects within the collective pinning volume $V_c \simeq \xi^2 L_c$.

To study the quantum  tunneling of the pancake vortex at a finite temperature,
we consider the path integral representation of the partition function
given by
\bena
Z(\beta \hbar)=\oint D[{\bf u }(\tau)] \exp(-S_E /\hbar),
\label{part}
\eena
where $\beta=1/k_B T$ and $S_E$ is the Euclidean action.
The path sum includes all the periodic paths 
${\bf u}(\tau)={\bf u} (\tau+\beta \hbar)$, where
${\bf u}$ is the displacement vector of the vortex  in the
$xy$ plane. The Euclidean action $S_E$
includes the Euclidean version of the Lagrangian $L_E$ :
\bena
S_E[{\bf u}(\tau)]=\int^{\beta \hbar}_{0} d\tau L_E [{\bf u}(\tau)].
\eena
The tunneling rate $\Gamma$  in the semiclassical limit,
with an exponential accuracy, is given by\cite{aff}
\bena
\Gamma \propto \exp[-S^{\rm min}_{E}(T)/\hbar].
\eena
We study $S^{\rm min}_E(T)$ which gives
the trajectory with the period
$\beta \hbar$ that minimizes the Euclidean action.\cite{wei}
Considering the situation where
the inertia term is not relevant and the vortex dynamics is
dominated by the Hall and the dissipative forces, we write the Euclidean
action as
\bena
S_E&=&  \int^{\beta \hbar}_{0} d\tau \{L_c \left[ -i \alpha { d u_x \over d \tau}
u_y + V(u_x, u_y) \right]
\nonumber \\
&&+\sum_{k} \left[ {1 \over 2} m_k ({\dot{x}}^{2}_{k}+
{\dot{y}}^{2}_{k} )+{1 \over 2} m_k \omega^{2}_{k}
\left(   \left( x_k-{C_k \over m_k \omega^{2}_{k}} u_x \right)^2+
\left(  y_k-{C_k \over m_k \omega^{2}_{k}} u_y \right)^2  \right) \right]  \},
\label{se1}
\eena
where $\alpha$ is the Hall coefficient and $V(u_x,u_y)$ is the pinning potential 
per unit length which includes the contribution from the Lorentz force.
The last term of Eq. (\ref{se1}) represents the dissipative
environment of the vortex consisting of
a set of harmonic oscillators as formulated 
by Caldeira and Leggett.\cite{cal}
The effect of the  dissipative environment is characterized by the
spectral function
\bena
J(\omega)={\pi \over 2} \sum_{k} {C^{2}_k \over m_k \omega_{k}}
\delta (\omega-\omega_k ).
\eena
With the oscillators integrated out,  the Euclidean action takes the form
\bena
S_E &=&  \int^{\beta \hbar}_{0} d\tau \{ L_c \left[ -i \alpha { d u_x \over d \tau}
u_y + V(u_x, u_y) \right]
\nonumber \\
&&+{1 \over  2} \int^{\infty}_{-\infty} d \tau^{\prime} K_0(\tau-\tau^{\prime})
\left[ (u_x (\tau)-u_x (\tau^{\prime}) )^2 +(u_y (\tau)- u_y (\tau^{\prime}) )^2 \right] \},
\label{act1}
\eena
where the nonlocal influence function is expressed as
\bena
K_0(\tau)={1 \over 2 \pi} \int^{\infty}_{0} d \omega 
J(\omega) \exp (-\omega |\tau| ).
\eena

\section{quantum tunneling of a vortex}

In order to study the motion of a vortex, we need to first
analyze the structure of the model potential 
$V(u_x, u_y)$. 
Since the external current  $j$ along the $y$ direction
brings the system into a metastable state 
by tilting the potential, the vortex has a chance to move
out of the  pinning potential. 
Let us define $u_{x_i}$ as the critical position of the vortex
at which the barrier vanishes at the critical
current $j_c$.    In the limit $j \rightarrow j_c$, 
$u_{x_i}$ and $j_c$ satisfy
\bena
\left[ {\partial V \over \partial u_{x} } \right]_{u_x=u_{x_i}}
=\left[ {\partial^2 V \over \partial u^{2}_{x} } \right]_{u_x=u_{x_i}}=0.
\eena
With $V(u_x,u_y)=V_p(u_x, u_y)-\phi_0 j u_x /c$,  $u_{x_i}$ and $j_c$
are given by the relations 
$(\partial V_p/\partial u_x )_{u_x=u_{x_i}} =\phi_0 j_c /c$
and $ (\partial^2 V_p/\partial u^{2}_{x} )_{u_x=u_{x_i}} =0$,
where $\phi_0=hc/2e$ is the flux quantum.
For the pinning potential, we choose an appropriate model potential
describing a typical tunneling situation: 
$V_p(u_x, u_y)$ should exhibit a local minimum and
should be connected via a saddle to the  free space along one direction (we 
choose this direction as the $x$ axis.).  A model potential 
satisfying the requirement is
\bena
V(u_x, u_y) \simeq {1 \over 2} V_0 \left[  
c_1 \epsn \left( { u_x \over R} \right)^2 - 
{2 \over 3} c_2  \left( { u_x \over R }  \right)^3  
+ \left( {u_y \over R}  \right)^2  \right],
\label{vpot}
\eena
where $c_1=R^2 \sqrt{2 \phi_0 j_c/(c V^{2}_{0})}
|(\partial^3 V /\partial u^{3}_{x} )_{u_x=u_{x_i}}|^{1/2} $,
$c_2 =R^3 |(\partial^3 V /\partial u^{3}_{x} )_{u_x=u_{x_i}}| /(2 V_0)$,  and
$\epsn= \sqrt{1-j/j_c} \ll 1$.
In Eq. (\ref{vpot}), $V_0$ and $R$ are
the height and the range of the pinning potential, respectively, and for
a typical weak pinning potential
$V_0  \simeq ( \phi_0/4 \pi \lambda_{xy} )^2 $ and $R (\geq \xi)$.
And $c_{1,2}$ are the dimensionless coefficients of the order of 1 and
$\lambda_{xy}$ is the bulk-planar penetration depth.

To consider  the tunneling of a vortex in the two regimes,
we investigate  the behavior of the Euclidean action (\ref{act1}).
In order to estimate  the order of magnitude of each term in the
action and to simplify the calculation for $\epsn \ll1$,
we introduce  the dimensionless  variables
\bena
\bar{u}_x=({ 2 c_2  \over  c_1 \epsn R }) u_x, \ \ \
\bar{u}_y=({2 c_2  \over  c_{1}^{3/2} \epsn^{3/2} R }) u_y, \ \ \
\bar{\tau}=({ \sqrt{c_1 \epsn} V_0  \over    \sqrt{2} R^2 \alpha_0  }) \tau,
\eena
where $\alpha_0=\pi \hbar n_s$ and $n_s$ is the number density of the
electrons in the condensate.

Assuming the Ohmic dissipation  where the frictional force acting on the
vortex is linear to the vortex velocity,\cite{bla94,bar,kop} 
the spectral density becomes
$J(\omega)= \eta \omega$, where
$\eta =(\pi/2) \sum_i (C^{2}_{i}/m_i \omega^{2}_{i}) 
\delta(\omega-\omega_i)$ =constant.\cite{cal}
With this choice, we have the influence function 
\bena
K_0 (\tau)={\eta \over 2 \pi |\tau|^2 },
\eena
which leads to the Euclidean action
\begin{eqnarray}
 S_E &=&
      ({\sqrt{2} c_{1}^{5/2}\over 4 c_{2}^{2}})
      (L_c \alpha_0 R^2)
      \epsn^{5/2}  I_{\rm HD},
      \label{se}
\end{eqnarray}
where
\begin{eqnarray}
 I_{\rm HD}&=&\int_{0}^{\Lambda} d\bar{\tau}
     \{ -i\alpha_1 {d\ubar_x \over d\bar{\tau}}
     \ubar_y +
     {1 \over 2}\ubar_{y}^{2}+
     {1 \over 2}\ubar_{x}^{2}-{1 \over 6}\ubar_{x}^{3}
     \nonumber \\
    & & + {1 \over 4} \eta_1
     \int_{-\infty}^{\infty}d\bar{\tau}_1
     {[\ubar_x(\taubar)-\ubar_x(\taubar_1)]^2+
     c_1 \epsn [\ubar_y(\taubar)-\ubar_y(\taubar_1)]^2
     \over |\taubar-\taubar_1|^2} \},
\label{IHD}
\end{eqnarray}
where  $\Lambda=\beta \hbar V_0 \sqrt{\epsn c_1}/(\sqrt{2} \alpha_0 R^2)$.
The dimensionless Hall  ($\equiv \alpha / (\sqrt{2} \alpha_0)$)
and dissipation coefficients  ($\equiv \eta / (L_c \sqrt{2\pi^2 c_1 \epsn} \alpha_0)$)
are given by\cite{kop}
\bena
\alpha_1  
&=& {1 \over \sqrt{2}}
{(\omega_0 \tau_r)^2 \over 1+(\omega_0 \tau_r)^2}, 
\label{a1} \\
\eta_1 
&=&{\sqrt{2} \over 2 \pi \sqrt{c_1 \epsn}}
 {\omega_0 \tau_r \over 1+(\omega_0 \tau_r)^2}.
\label{eta1}
\eena
Here, $\omega_0$ is the
level spacing of the quasiparticle bound states inside
the vortex core and $\tau_r (=m/n e^2 \rho_n)$ is a quasiparticle scattering time,
where $n$ is the number density of the charge carriers and $m$ and  $\rho_n$ are
their effective mass and resistivity, respectively. 
As can be seen in Eqs. (\ref{a1}) and (\ref{eta1}),
the Hall coefficient $\alpha$ is reduced from its pure value $\pi \hbar n_s$ due
to the dissipative effect. Although Ao {\it et al.} suggested that
the Hall coefficient is originated from the topological property and
thus not renomalized,\cite{ao} it seems that at least some aspects of
the experimental behavior\cite{dal} can be understood on the basis of
the renomalization of the Hall coefficient. Therefore, it is
meaningful to take $\alpha$ and $\eta$ to be
two parameters determined by the magnitude 
of $\omega_0 \tau_r$.

\subsection{Action in the Fourier Space}

When the Hall and the dissipative dynamics
are simultaneously present, the classical trajectories of $\ubarx$ and $\ubary$
satisfy 
\begin{eqnarray}
 i\alpha_1 {d \ubary \over d\taubar}+\ubarx
-{\bar{u}_{x}^{2} \over 2}
 -\eta_1 \int_{-\infty}^{\infty}d\taubar_1 
 ({d\ubarx \over d \taubar_1}) 
{1 \over \taubar_1-\taubar}&=&0,
 \label{eq:h-d-t1}\\
 -i\alpha_1 {d\ubarx \over d\taubar}+\ubary
 -\eta_1 c_1 \epsilon \int_{-\infty}^{\infty}d\bar{\tau}_1
 ({d\ubary \over d \taubar_1})
{1 \over \taubar_1-\taubar}&=&0.
 \label{eq:h-d-t2}
\end{eqnarray}
The substitution $\taubar \rightarrow -\taubar$ in Eqs. (\ref{eq:h-d-t1}) and
(\ref{eq:h-d-t2}) shows the invariance of the equations by taking
$\ubarx(-\taubar)=\ubarx(\taubar)$
and $\ubary(-\taubar)=-\ubary(\taubar)$. 
We will keep $c_1$ in the ensuing equations, although 
we will take $c_1= 1$ for the
numerical calculations.
Denoting $\ubarbf(\taubar) \equiv (\ubarxt, \ubaryt)$, we have
$\ubarbf(\taubar+\Lam)=\ubarbf(\taubar)$ at a finite temperature.
A simple analysis shows that $\ubarxt$ is real and $\ubaryt$ pure
imaginary, so they can be expanded into the Fourier series:
\bena
\ubarxt=\sum^{\infty}_{n=-\infty} u_n \exp (i \ombarn \taubar), 
\label{ubarx} \\
\ubaryt=-i \sum^{\infty}_{n=-\infty} v_n \exp (i \ombarn \taubar),
\label{ubary}
\eena
where $\ombarn=2 \pi n/\Lam$ (n=0, 1, 2...).
Substituting them into Eqs. (\ref{eq:h-d-t1}) and (\ref{eq:h-d-t2}),
we have
\bena
\left( 1+\pi \eta_1 |\ombarn|+
{ \alpha^{2}_{1} \bar{\omega}^{2}_{n} \over 1+\pi \epsn \eta_1 c_1 |\ombarn|}
\right)
u_n={1 \over 2} \sum^{\infty}_{m=-\infty} u_{n+m} u_m,
\label{un} \\
v_n =-{ i \alpha_1 \bar{\omega}_{n} \over 1+\pi \epsn \eta_1 c_1 |\ombarn|} u_n,
\label{vn}
\eena
where $u_{-n}=u_n$ and $v_{-n}=-v_n$.
Although Eq. (\ref{un}) is a one-dimensional problem with respect to $u_n$,
its solution becomes complicated  by the presence of the nonlocal term arising
from the cubic potential. For general $\omega_0 \tau_r$ and $\epsn$, we have
numerically solved Eqs. (\ref{un}) and (\ref{vn}) via the variational method.
The trial function for the variational method has been taken by combination of the
analytic solutions in the two extreme limits as follows.
At zero temperature,
$\ubarx (\bar{\omega})$'s are of the form
$\bar{\omega}/\sinh( \alpha_1 \pi \bar{\omega})$ in the Hall limit 
($\omz \tau_r \rightarrow \infty$) and $\exp(-\pi \eta_1 |\bar{\omega}|)$
in the dissipative limit ($\omz \tau_r \rightarrow 0$),\cite{kim98} so
a natural choice 
for the trial function at finite temperatures is 
\bena
u_n={p_1 \ombarn \over \sinh(p_2 \ombarn) }+p_3 \exp(-p_4 |\ombarn|),
\eena
where $p_i$'s ($i$=1,2,3,4)  are free parameters to be determined
by the variational method. It turns out that the numerical variational
method with the   trial function works very successfully.

Using the Fourier series in Eqs. (\ref{ubarx}) and  (\ref{ubary}), we write
$I_{\rm HD}$ as
\bena
\IHD=\Lam \suminf
\left[ -\alpha_1 \ombarn u_n v_n +{1 \over 2} u^{2}_{n}-{1 \over 2} v^{2}_{n}
-{1 \over 6} u_{n} \left( \sum^{\infty}_{m=-\infty}   u_{n+m} u_m  \right) + 
{\pi \over 2} \eta_1 |\ombarn|
(u^{2}_{n}- c_1 \epsn v^{2}_{n} ) \right],
\eena
which further reduces to
\bena
\IHD={1 \over 6} \Lam \suminf \left( 1+\pi \eta_1 |\ombarn|+
{ \alpha^{2}_{1} \bar{\omega}^{2}_{n} \over 1+ \pi \epsn \eta_1 c_1 |\ombarn| }
\right) u^{2}_{n}.
\label{IHDf}
\eena

\subsection{Quantum Relaxation  near the 
Crossover Temperature and at Zero Temperature}

At $T_c$, the crossover temperature between thermal activation and
quantum tunneling,
the classical trajectories become independent of
$\bar{\tau}$,  i.e., $\ubarxt=2$ and $\ubaryt=0$. When
$\Lam(T)$ is slightly greater than  $\Lam_c [\equiv \Lambda(T_c) ]$,
we take only the first Fourier harmonics for the solution
because the next harmonics are smaller near $T_c$:
\bena
\ubarxt&=&u_0+2 u_1 \cos \left({2 \pi \over \Lam} \taubar \right), \\
\ubaryt&=& 2 v_1 \sin \left({2 \pi \over \Lam} \taubar \right).
\eena
Exploiting the fact that $u_n$'s are zero except for $u_0$ and
$u_1$ in Eq. (\ref{un}), we get
\bena
u_0 &=&1+{2 \pi^2 \eta_1 \over \Lam}+ { 4 \pi^2 \alpha^{2}_{1}
\over \Lam(\Lam+2 \pi^2 c_1 \epsn \eta_1)}, 
\label{u0} \\
u^{2}_{1}&=&u_0-{1 \over 2} u^{2}_{0}.
\label{u1}
\eena
Setting $u_0=2$ in Eq. (\ref{u0}) and solving for $\Lam=\Lam_c$, we have
\bena
\Lam_c={4 \pi (\alpha^{2}_{1}+\pi^2 \eta^{2}_{1} c_1 \epsn ) \over
\sqrt{ \pi^2 \eta^{2}_{1} (1+c_1 \epsn)^2+4 \alpha^{2}_{1} } -\pi \eta_1 (1-c_1 \epsn) }.
\label{lamc}
\eena
Using the relations in Eqs. (\ref{a1}) and (\ref{eta1}), we plot $\Lam_c$
against $\omz \tau_r$ for different values of  $\epsn$  in
Fig. \ref{fig:crossover}.
The maximal values of the crossover  periods are more
pronounced in the limit of smaller  $\epsn$, and $\Lam_c$'s converge
to $2 \pi \alpha_1 (=\sqrt{2} \pi)$   in the Hall regime and to
$2 \pi^{2}\eta_1 (=\sqrt{2} \pi \omztau/\sqrt{\epsn})$ 
in the dissipative regime. The reduced action
integration near the crossover temperature can also
be simply obtained by summing only
$n=0$ and $n=1$ contributions:
\bena
\IHD={1 \over 6} \Lam \left[ u^{2}_{0}+2
\left( 1+{2 \pi^2 \eta_1 \over \Lam}+ { 4 \pi^2 \alpha^{2}_{1}
\over \Lam(\Lam+2 \pi^2 c_1 \epsn \eta_1) } \right) u^{2}_{1}   \right],
\eena
which is reduced to
\bena
\IHD={1 \over 6} \Lam u^{2}_{0} (3- u_0 ),
\label{IHD-cross}
\eena
by using Eqs. (\ref{u0}) and (\ref{u1}).

The action integration $I_{\rm HD}$ obtained by the numerical variational
method around $T_c$ and the one by Eq. (\ref{IHD-cross}) are
compared in Fig. \ref{fig:qtq0}.
The two curves in the figure perfectly
join at the crossover period, which implies that
our numerical method gives the correct solution.
The dynamical magnetization relaxation rate $Q$  is given by 
$Q=\hbar/S_E$.
In real experiments   $Q$ is
extracted from the magnetization $M(t)=M_0 [1-Q \ln (t/t_0)]$\cite{yes88}.
From Eq. (\ref{se}), we have
$Q(T)/Q_0=2 \sqrt{2}/I_{\rm HD}$, where
$Q_0=(\pi n_s L_c R^2\epsn^{5/2})^{-1}$ by taking $c_1=c_2=1$.
Then, at the crossover temperature $T_c$, since
$I_{\rm HD}=2 \Lam_c/3$, we have
\bena
{Q(T_c) \over Q_0 }={3 \sqrt{2} \over \Lam_c}.
\label{qtc}
\eena
In Fig. \ref{fig:act-cross}, $Q(T_c)/Q_0$ and $Q(0)/Q_0$ are plotted with respect
to $\omz \tau_r$. It  is interesting 
that  the shape of  $Q(T_c)/Q_0$
for each $\epsn$ is close to that of $Q(0)/Q_0$ at
zero temperature\cite{kim98}. This fact can be understood by considering
the following features of the relaxation rate. 
Since the tunneling rate $\Gamma$ is expected to be almost temperature 
independent   for $T \leq T_c$
and $\Gamma \sim \exp(-U/k_B T)$ for $T \geq T_c$, the 
crossover temperature is approximately given
by the relationship $U/(k_B T_c) \simeq S_E (T=0)/\hbar $. And
$Q(0)=\hbar/S_E(T=0) \simeq (k_B T_c) / U=Q(T_c)$.
Hence, we obtain the
analytic expression for the relaxation
rate at zero temperature using Eqs. (\ref{lamc}) and (\ref{qtc})
\bena
{Q(0) \over Q_0} \simeq {3 \over 2\pi}
{\sqrt{(1+\epsn)^2 +4 (\omz \tau_r)^2 \epsn} -(1-\epsn) 
\over (\omz \tau_r)\sqrt{\epsn} },
\label{qtzq0}
\eena
by taking $c_1=1$ and using 
$\alpha_1/\eta_1=\omz \tau_r \pi\sqrt{\epsn}$.
Eq. (\ref{qtzq0}) agrees with the result in Ref. \cite{mel} up to a
numerical factor.
Since the analytic form for $Q(0)/Q_0$ is known to be 5/6 in the limit
of $\omztau \rightarrow \infty$,\cite{kim98} the correct
prefactor of Eq. (\ref{qtzq0}) is  5/12 instead of  $3/(2\pi)$.
Although $Q(0)/Q_0$ goes to  infinity as
$\omega_0\tau_r \rightarrow 0$ in Eq. (\ref{qtzq0}), 
it actually  does not diverge in that
limit, because, by including the inertia term not considered in this work, 
the approximate form of the classical action in the limit becomes
$S_E/\hbar \sim L_c \sqrt{m_v V_0} \xi 
\epsn^{5/2}$, which is independent of $\omega_0\tau_r$,  where
$m_v$ is the inertia mass of a vortex.\cite{bla94,suh,dua}
In general, the mass term is relatively small in the
Hall and dissipative regime and can usually be neglected.  

In a  moderately clean regime, for small values of $\epsn$
each curve in Fig. \ref{fig:act-cross} has a minimum  around 
$\omztau = 1$, which is interested.
In fact, from Eq. (\ref{qtzq0}) we can see that the position of the minimum 
at zero temperature is
$\omztau =(1+\epsn)/(1-\epsn)$.
As $\epsilon$ becomes smaller, i.e., as $j \rightarrow
j_c$, the minimum becomes much more pronounced with its location
moving toward $\omega_0\tau_r = 1$ at the same time.  
The existence of such minima can be  understood by considering the 
following qualitative
features of the relaxation rates in the two regimes.
Since $\ubarx (\omega)$ is proportional
to $ \alpha_1 \omega/\sinh(\alpha_1 \pi \omega)$  in the Hall limit,\cite{kim98}
the classical trajectories with $|\omega| \lesssim 1/\alpha_1$
contribute to the Euclidean action mostly.
From Eqs. (\ref{se}) and (\ref{IHDf})
the correction to the Hall action $S^{(H)}_{E}$ by the small dissipation is given by
$(1+\eta_1/\alpha_1)S^{(H)}_{E} \sim (1+1/\omega_0 \tau_r) S^{(H)}_{E}$, which 
leads to the relaxation rate given by
$Q(0) \sim \omega_0 \tau_r/(1+\omega_0 \tau_r)$.
So the relaxation rate $Q$ decreases with decrease in
$\omega_0 \tau_r$ 
from $\infty$, which  is also physically clear because
the classical action increases by inclusion
of the dissipation.
In the opposite limit  the correction to 
the purely dissipative action $S^{(D)}_{E}$ by the small Hall contribution
is $[1+(\alpha_1/\eta_1)^{2}] S^{(D)}_{E} 
\sim \omega_0 \tau_r [1+(\omega_0 \tau_r)^2 ]$, leading to the 
relaxation rate $Q(0) \sim 1/[\omega_0 \tau_r (1+(\omega_0 \tau_r)^2)]$.
In this case,
$Q(0)$ decreases with increase in $\omega_0 \tau_r$.
Therefore, a minimum in $Q(0)$ should exist
in the intermediate regime, which 
suggests  the existence of the strong pinning in the 
moderately clean samples.

\subsection{Quantum Relaxation in  Dissipation Regime}

In the dissipative limit, we take $\alpha_1 \rightarrow 0$
in the action integration of Eq. (\ref{IHDf}). The reduced action then becomes
\bena
\IHD={1 \over 6} \Lam \suminf \left( 1+\pi \eta_1 |\ombarn|
\right) u^{2}_{n} \equiv I_D.
\eena
Noting that $u_n$ in the dissipative limit is given by
$u_n=u_0 \exp(-b |n|)$ where $u_0=4 \pi^2 \eta_1/\Lam$ and
$b=\tanh^{-1}(2 \pi^2 \eta_1/\Lam)$,\cite{lar}
we get the reduced action given by
\bena
I_{D}=\Lam_c  \left[ 1-{1 \over 3} \left( {T \over T_c} \right)^2  \right],
\label{I-diss}
\eena
where $\Lam_c=2 \pi^2 \eta_1$ and 
$k_B T_c=\hbar \sqrt{c_1 \epsn} V_0 /(2\sqrt{2} \pi^2 R^2 \alpha_0 \eta_1)$.
In Fig. \ref{fig:act-diss}, the
relaxation rate using Eq. (\ref{I-diss}) is compared with the one
obtained from the
numerical solution: the two
curves match quite well asymptotically in the region of small
$\omz \tau_r$ values.

\subsection{Quantum Relaxation in  Hall Regime}

In the Hall limit we take $\eta_1 \rightarrow 0$ in Eq. (\ref{IHDf}), which
leads to the reduced action given by
\bena
\IHD={1 \over 6} \Lam \suminf \left( 1+
\alpha^{2}_{1} \bar{\omega}^{2}_{n} 
\right) u^{2}_{n} \equiv I_H,
\label{IHf}
\eena
where $u_n$ satisfies
\bena
(1+\alpha^{2}_{1} \bar{\omega}^{2}_{n}) u_n={1 \over 2} \suminf u_{n+m} u_m.
\label{unH}
\eena
While  the  instanton solution can be obtained analytically in the dissipative regime,
the solution of Eq. (\ref{unH}) can be found numerically.
We use Eqs. (\ref{eq:h-d-t1}) and  
(\ref{eq:h-d-t2}) rather than Eq. (\ref{unH}) and  obtain the reduced differential
equation 
\bena
2 \alpha^{2}_{1} {d^2 \bar{u}_x \over d \bar{\tau}^2 } -2 \ubarx + \bar{u}^{2}_{x}=0.
\label{uh}
\eena
We then integrate for $I_{\rm HD}$ with $\eta_1=0$ in Eq. (\ref{IHD})
using the solution for $\ubarx$ in Eq. (\ref{uh}) and obtain 
the reduced action $I_H$.
As in the case of the dissipative regime, we have found that
the  relaxation rate 
agrees  with the one obtained from the variational procedure in the limit of 
$\omz \tau_r \rightarrow \infty$.

\section{Discussion  at finite temperature}

We now consider the problem at the finite temperature  in the whole  regime, 
i.e., when the
Hall and the dissipative dynamics are simultaneously present.
The solutions for $\bar{u}_x (\tau)$ and $\bar{u}_y (\tau)$ in Eqs.
(\ref{ubarx}) and (\ref{ubary}) are obtained through
$u_n$'s and $v_n$'s of Eqs. (\ref{un}) and (\ref{vn}) which
are numerically obtained by the variational method. 
In Fig. {\ref{fig:inst}, we show
$\ubarxt$ and $\ubaryt$ for various periods $\Lam$
which exhibit the typical trend of the classical trajectories
as the period is successively shortened. The peak-to-valley amplitudes of
$\ubarxt$ and $\ubaryt$ decrease as the period gets shorter, eventually becomes
flat at $T_c$, i.e., $\ubarxt=2$ and $\ubaryt=0$. We subsequently
calculate the
reduced action (\ref{IHD}) via (\ref{IHDf}) and the corresponding 
relaxation rate.
The three-dimensional plot of $Q(T)/Q_0$ versus $\omz \tau_r$ and $\Lam$
for $\epsn=0.01$ is shown in Fig. \ref{fig:three-qtq0-e001}. 
We have also plotted $Q(T)/Q_0$ against $\Lam$ for the different values
of $\omz \tau_r$ in Fig. \ref{fig:qtq0-t-e001}. As can be seen in the figure, $\omz \tau_r=1$
is the boundary of the two different behaviors of the relaxation rate: for
$\omz \tau_r< 1$,
the relaxation rate increases with decreasing
$\omz \tau_r$, whereas for $\omz \tau_r >1$ 
it increases with increasing $\omz \tau_r$. 
The dependence of $Q(T)/Q_0$ on $\omz \tau_r $
is shown in Fig. \ref{fig:qtq0-om-e001}
for the different values of the period. 

What is interesting is the behavior of the relaxation rate in the
intermediate region of $\omztau$. 
We focus our attention on the four
different temperature regimes, as indicated in Fig.
\ref{fig:qtq0-om-e001} (b).
If the temperature is sufficiently low so that 
$\Lambda > \Lambda^{(m)}_{c}$, the line yields no
intersection points, and there exists quantum relaxation in the whole regimes of
$\omz \tau_r$. If the temperature is sufficiently high so that 
$\Lambda <\Lambda^{(H)}_{c}$, quantum relaxation occurs only in the dissipative
regime ($\omz \tau_r < (\omz \tau_r )_D$). 
In the temperature range $\Lambda^{(H)}_{c} < \Lambda
< \Lambda^{(m)}_{c}$,  on the other hand, quantum relaxation
exists either in the Hall regime or in the dissipative regime, and
purely thermal relaxation occurs in the crossover region between
the two regimes.
The values for $\Lambda^{(m)}_{c}$,
$\Lambda^{(H)}_{c}$, and $ (\omz \tau_r )_D$ can readily be computed
from the position of the minimum and using Eq. (\ref{lamc}) with $c_1=1$:
$\Lambda^{(m)}_{c}=\pi(1+\epsn)/\sqrt{2 \epsn}$, 
$\Lambda^{(H)}_{c}=\sqrt{2} \pi$, and
$(\omz \tau_r )_D=\sqrt{\epsn}/(1-\epsn)$.
The corresponding relaxation rates are given by
$Q(\Lambda^{(m)}_{c})/Q_0=6 \sqrt{\epsn}/[\pi(1+\epsn)]$ and
$Q(\Lambda^{(H)}_{c})/Q_0=3/\pi$.
The minimum of $\qtqz$ in the intermediate regime is then
noticed.
As the temperature becomes lower, the
quantum relaxation rate is
more developed in the intermediate regime, and at an extremely low
temperature it has a minimum at $\omztau \sim 1$.
This feature is more pronounced for smaller  $\epsn$ and larger $\Lam$.
Correspondingly, in such a regime
the quantum depinning of a vortex is expected to be smaller at  lower temperatures
in the regime.

Before concluding, we illustrate our results
with specific  numbers. In the experiment of Ref. \cite{dal}, the 
relaxation rate is $Q(0)/Q_0 \sim 2.3~ (2.0)$ in the YBCO (BiSCCO) system.
In this case,   $\omztau \sim 0.29~ (0.37)$ for YBCO (BiSCCO)
which corresponds to the Hall angle
$\Theta_H=\arctan(\omztau) \sim 16^\circ~ (20^\circ)$,
which depends on the oxygen content.
The numbers imply that the samples are moderately clean.
However, the regime which was considered in Ref. \cite{dal} was
$\omztau \lesssim 1$, where the onset of the minimum just takes place.
In order to observe the minima,
the experiment should be extended to the region $\omztau \gg 1$.

\section{Conclusions}

In conclusion, we have considered quantum tunneling of a vortex in the
presence of the Hall and the dissipative dynamics.
We have derived the analytic expression for the relaxation
rate at zero temperature and obtained the numerical solutions
by the variational method at finite temperatures.
The relaxation rate is constant in the Hall limit and proportional to
$1/(\omztau)$ in the dissipative limit, and, consequently, a minimum 
exist at  $\omztau=2(j_c/j)(1+\sqrt{1-j/j_c})-1$. Therefore, the strongest pinning is 
expected in the moderately
clean sample at zero temperature. 
At finite temperatures, the quantum relaxation rate tends to vanish in the
intermediate regime where both the Hall and the dissipative terms contribute
to the dynamics of a vortex. At sufficiently low temperatures,
quantum vortex tunneling occurs in the whole regime and the corresponding relaxation
rate has a minimum at  $\omz \tau_r \sim 1$.
These features are expected to be observed in future
experiments.

\acknowledgments

This work was supported 
by grant No. R01-1999-00026 from the Korea Science and Engineering Foundation.

\pagebreak

\begin{figure}
\caption{ $\Lam_c$ versus  
$\omz \tau_r$ where $c_1=1$.  Note that the crossover
temperatures become independent of $\epsn$  in the
Hall regime ($\omz \tau_r \rightarrow \infty$). }
\label{fig:crossover}
\end{figure}

\begin{figure}
\caption{ The  relaxation rate $Q(T)/Q_0$ versus $\Lam$ near the crossover temperature for
$\epsn=0.1$ and $\omz \tau_r$ =1,  where
$\Lam_c  \sim 7.6$. The solid line represents the 
analytical curve from Eq. (\protect\ref{IHD-cross}), and the dotted line with diamonds indicates
the result of the numerical calculation.
}
\label{fig:qtq0}
\end{figure}

\begin{figure}
\caption{ The relaxation rate evaluated at the crossover temperature:
$Q(T_c)/Q_0$ versus $\omz \tau_r$ 
where $\epsn=$ 0.1 (a), 0.01 (b), and 0.001 (c). Inset: 
the relaxation rate $Q(0)/Q_0$ at zero temperature
with $\epsn=$ 0.1 (a), 0.01 (b), and 0.001 (c).
}
\label{fig:act-cross}
\end{figure}

\begin{figure}
\caption{ The relaxation rate in the dissipative limit when $\epsn=0.1$ and
$\Lam=10$. The solid line represents the evaluation of Eq. (\protect\ref{I-diss}),
and the diamonds are the numerical results from the variational method.
}
\label{fig:act-diss}
\end{figure}

\begin{figure}
\caption{ Typical instanton solutions with different periods: $\ubarxt$ (top)
and $-i\ubaryt$ (bottom) for $\epsn=0.1$ and $\omz \tau_r=1.0$, where the
periods are $\infty$ (a), 10 (b), 8  (c), and 7.617 (d). 
}
\label{fig:inst}
\end{figure}

\begin{figure}
\caption{ The relaxation rate $Q(T)/Q_0$ for $\epsn=0.01$  against $\omz \tau_r$ and $\Lam$.
In order to show the curve for $Q(T_c)/Q_0$, we have omitted the
purely thermal relaxation rate. See Fig. \protect\ref{fig:qtq0-om-e001} for 
details.
}
\label{fig:three-qtq0-e001}
\end{figure}

\begin{figure}
\caption{ The relaxation rate $Q(T)/Q_0$ versus $\Lam$ for the different values
of $\omz \tau_r$ when $\epsn=0.01$. $\omz\tau_r$  increases
from the bottom ( $\omztau=1$) and approaches $\infty$. 
Inset: The case for $\omztau \leq 1$.
$\omz\tau_r$ decreases from the bottom.
}
\label{fig:qtq0-t-e001}
\end{figure}

\begin{figure}
\caption{ (a) $Q(T)/Q_0$ versus $\omz \tau_r$ for different 
periods when $\epsn=0.01$.
Periods are 30, 20, 15, and 10 (from the bottom).
The curve with the period of 30 is already very close to that of an
infinite period, which corresponds to zero temperature. 
The dotted curve  is $Q(T_c)/Q_0$,
of Fig. \protect\ref{fig:act-cross}. In the region above the 
dotted curve, purely
thermal relaxations exist along the horizontal lines.
(b) A schematic diagram of $\Lambda_c$ versus $\omztau$
with the  lines of constant temperatures. Four  
cases are considered : (I) $\Lambda > \Lambda^{(m)}_{c}$, 
(II) $\Lambda=\Lambda^{(m)}_{c}$, (III) 
$\Lambda^{(H)}_{c}<\Lambda < \Lambda^{(m)}_{c}$, and
(IV) $\Lambda \leq \Lambda^{(H)}_{c}$.
Note that $\Lambda^{(m)}_{c}=22.4$, $Q(T_c)/Q_0=0.189$, 
and $(\omztau)_D=0.101$. 
}
\label{fig:qtq0-om-e001}
\end{figure}

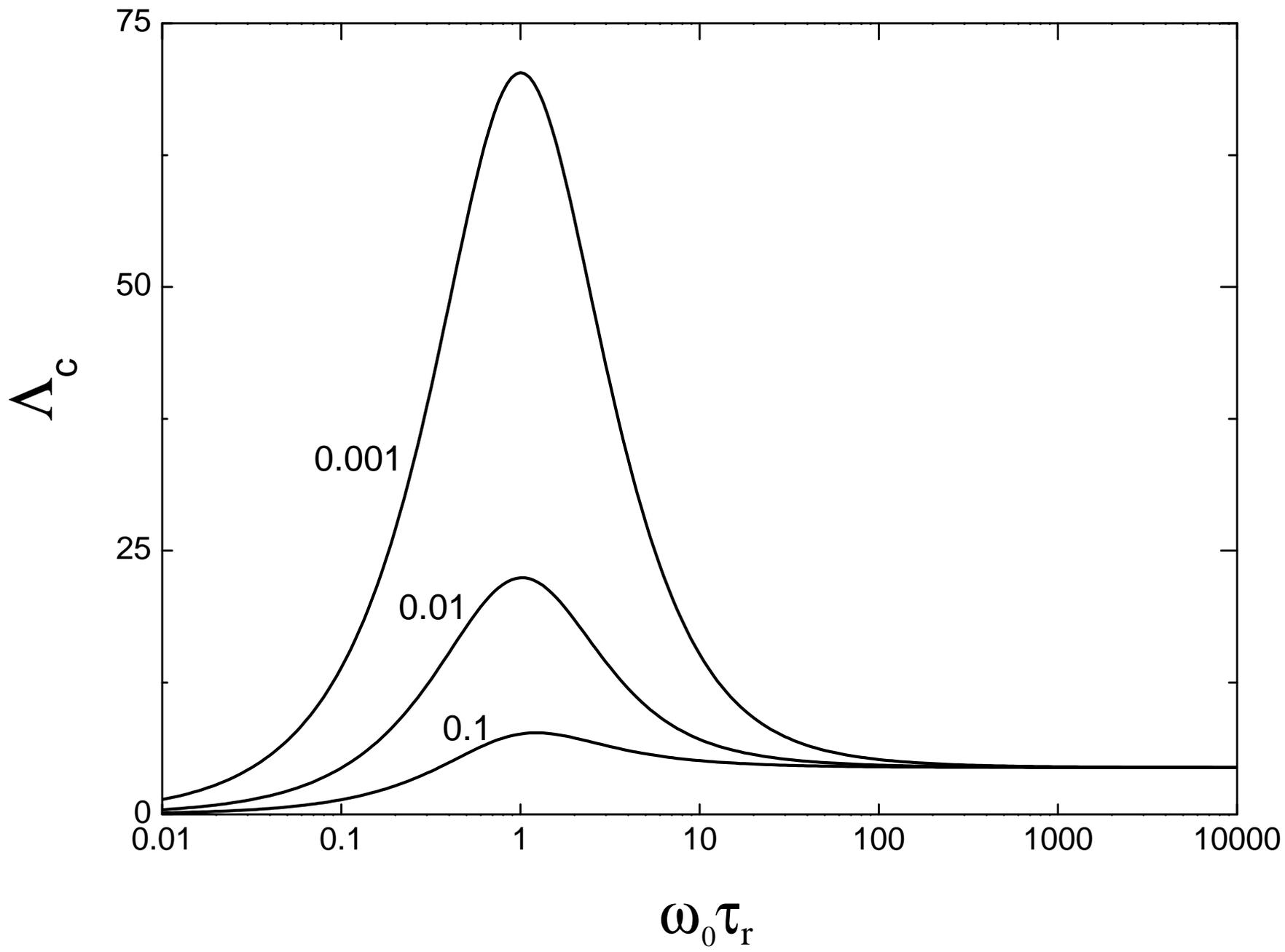
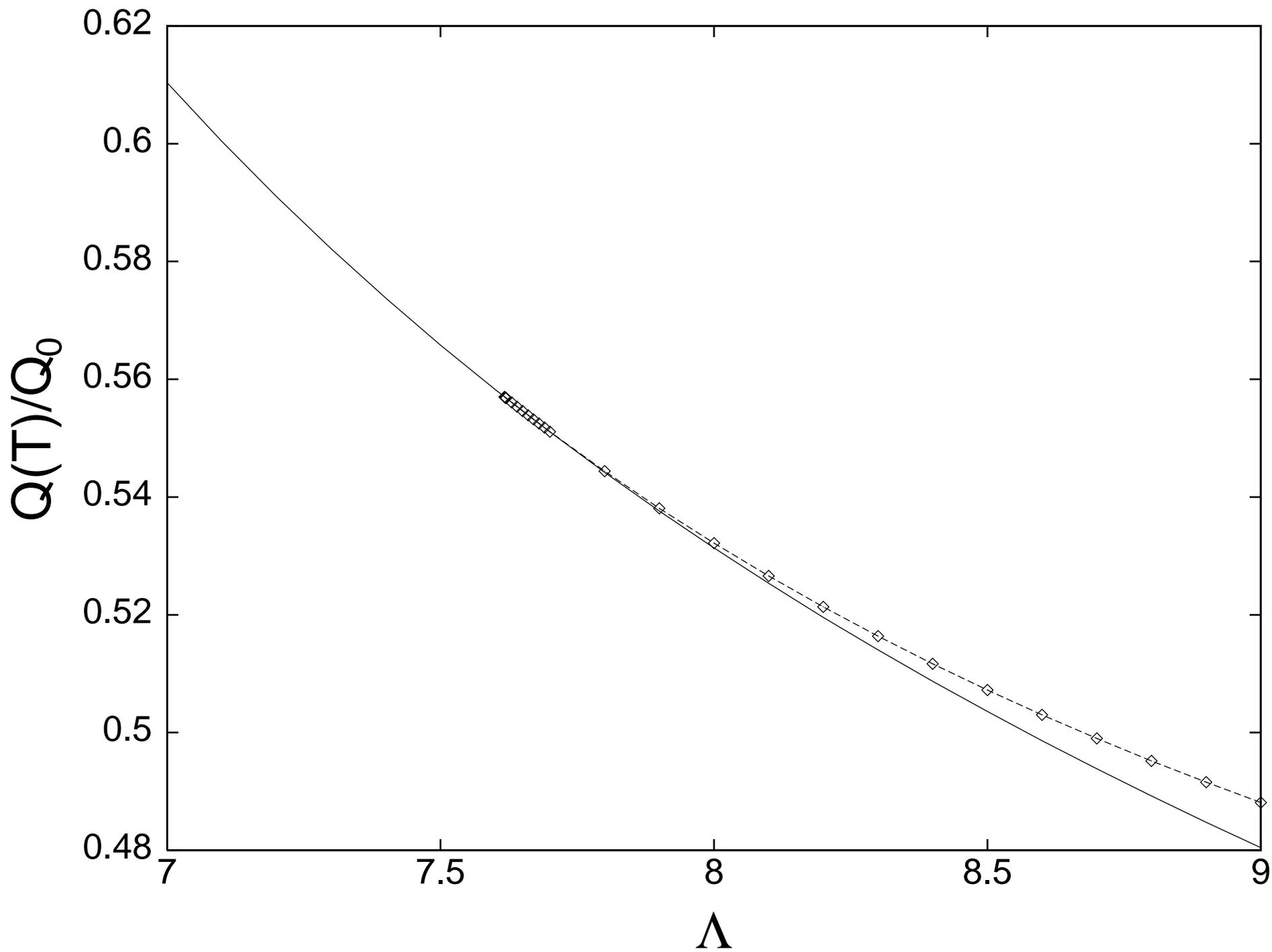
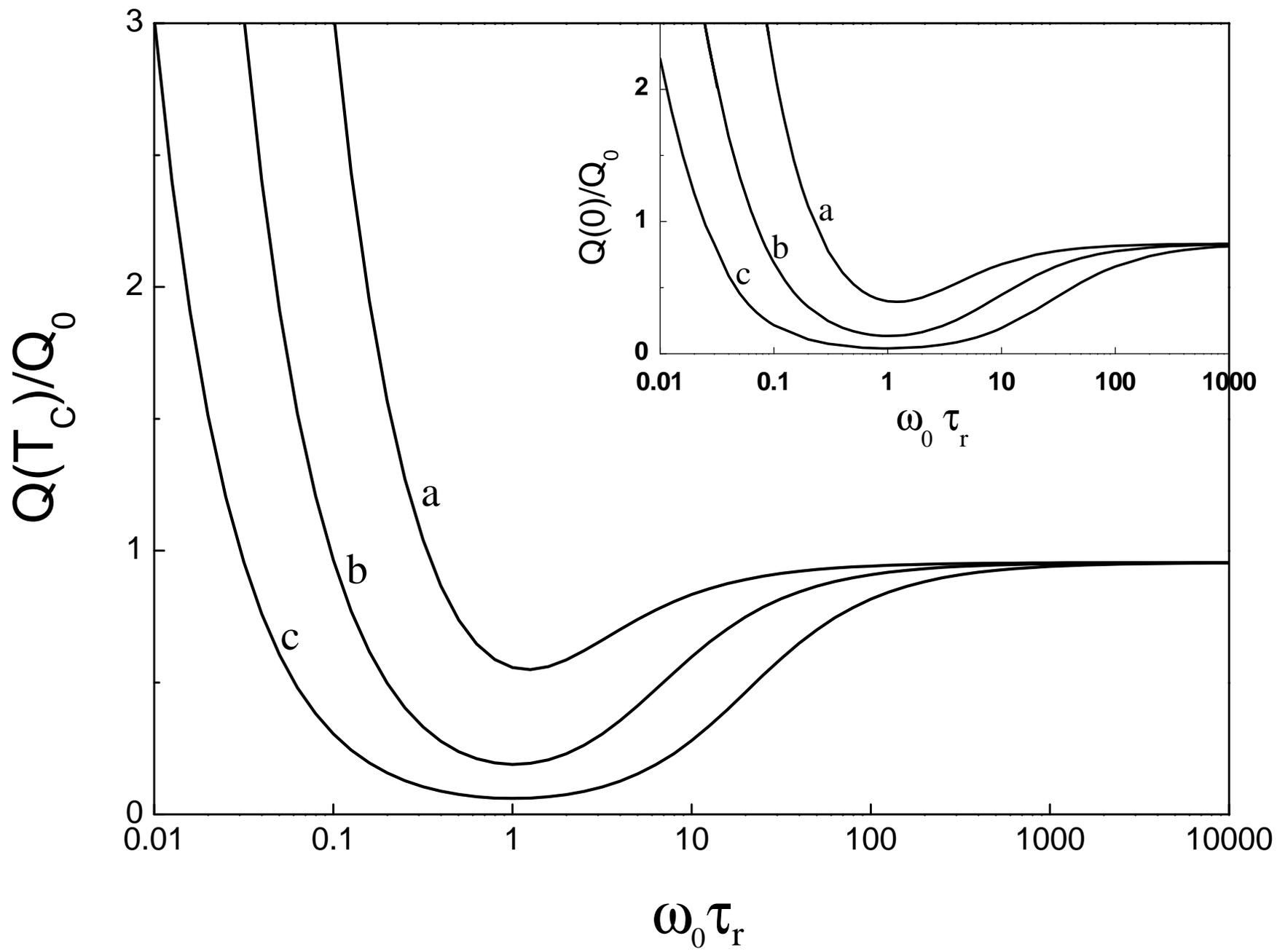
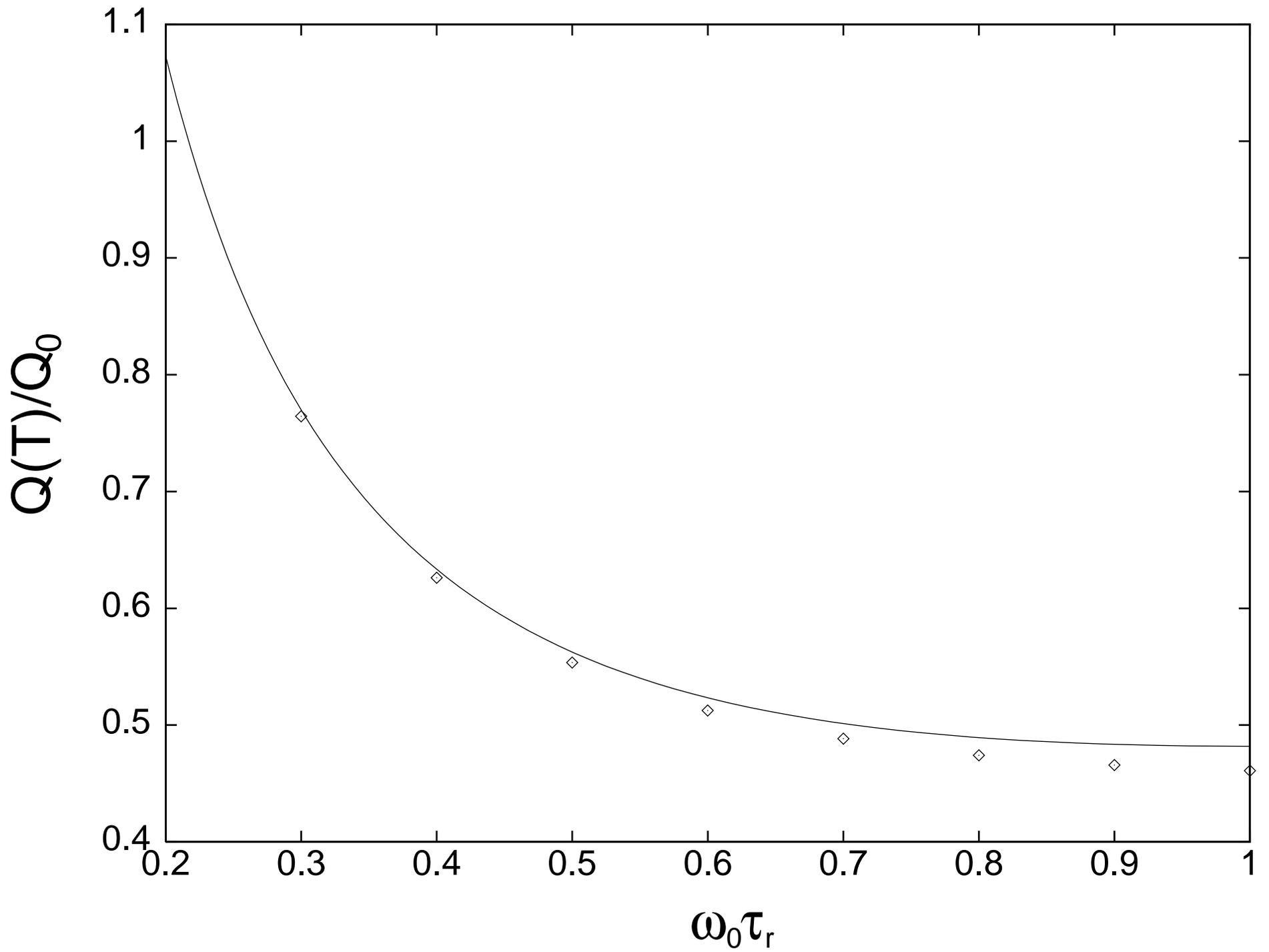
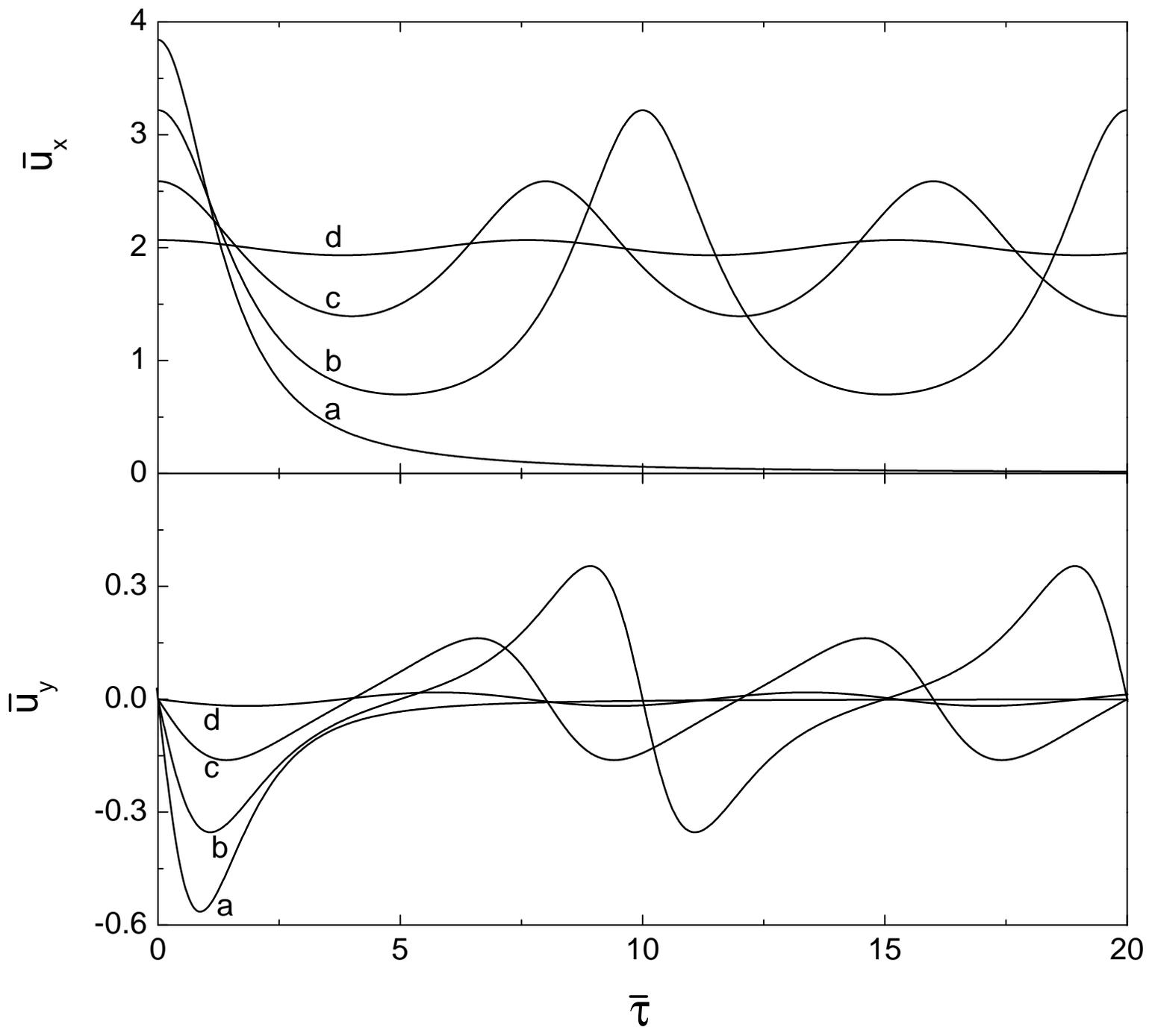
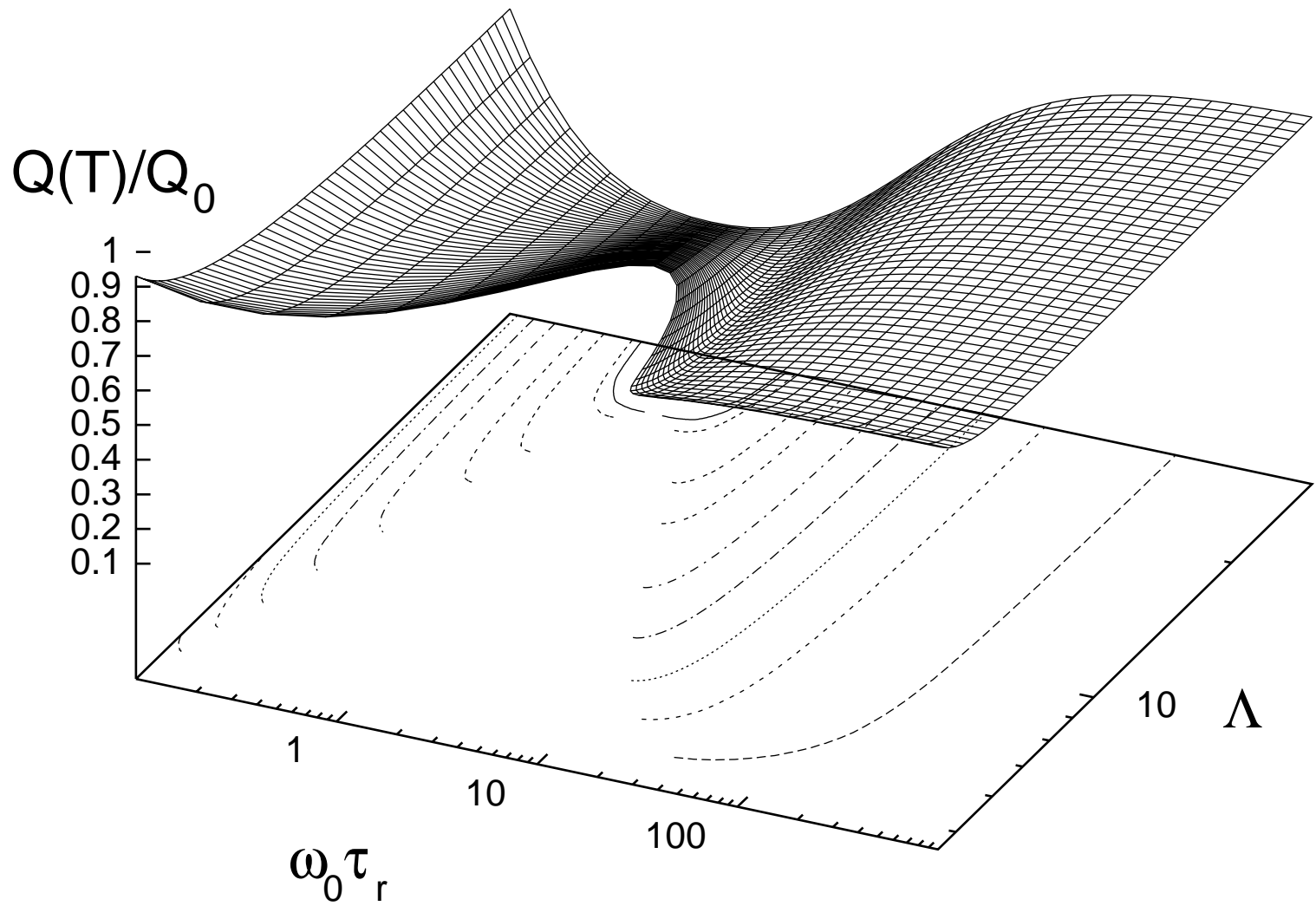
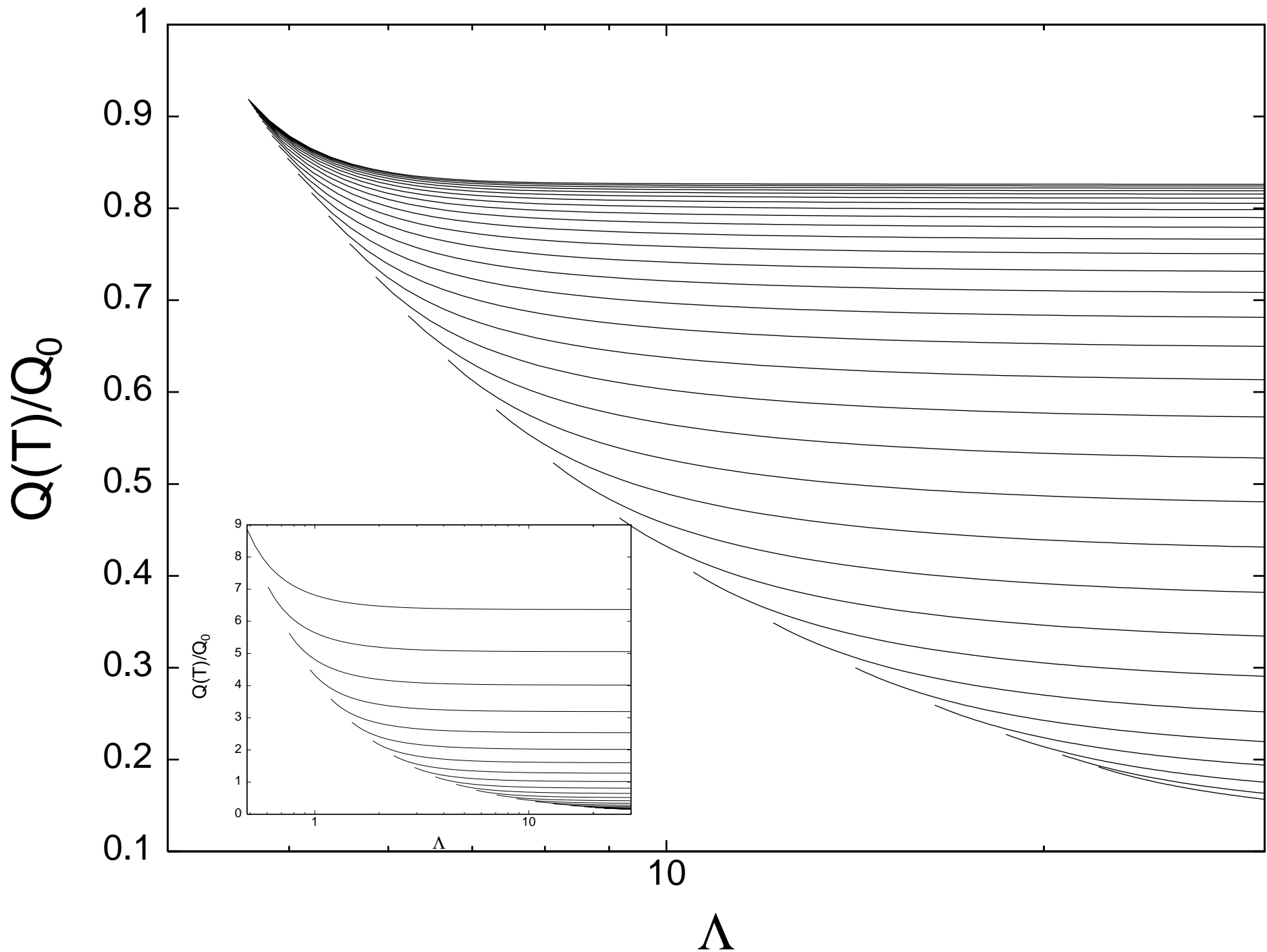
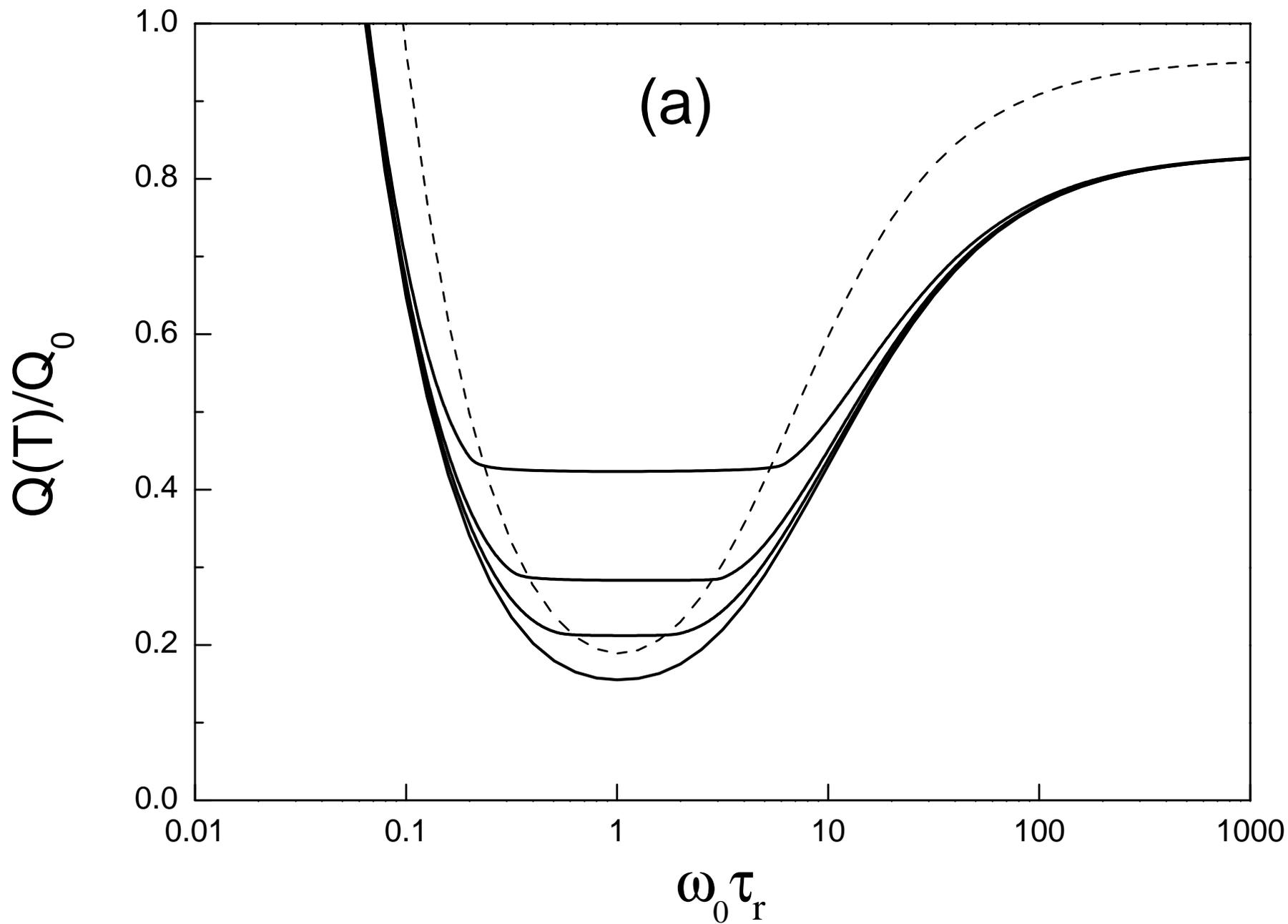
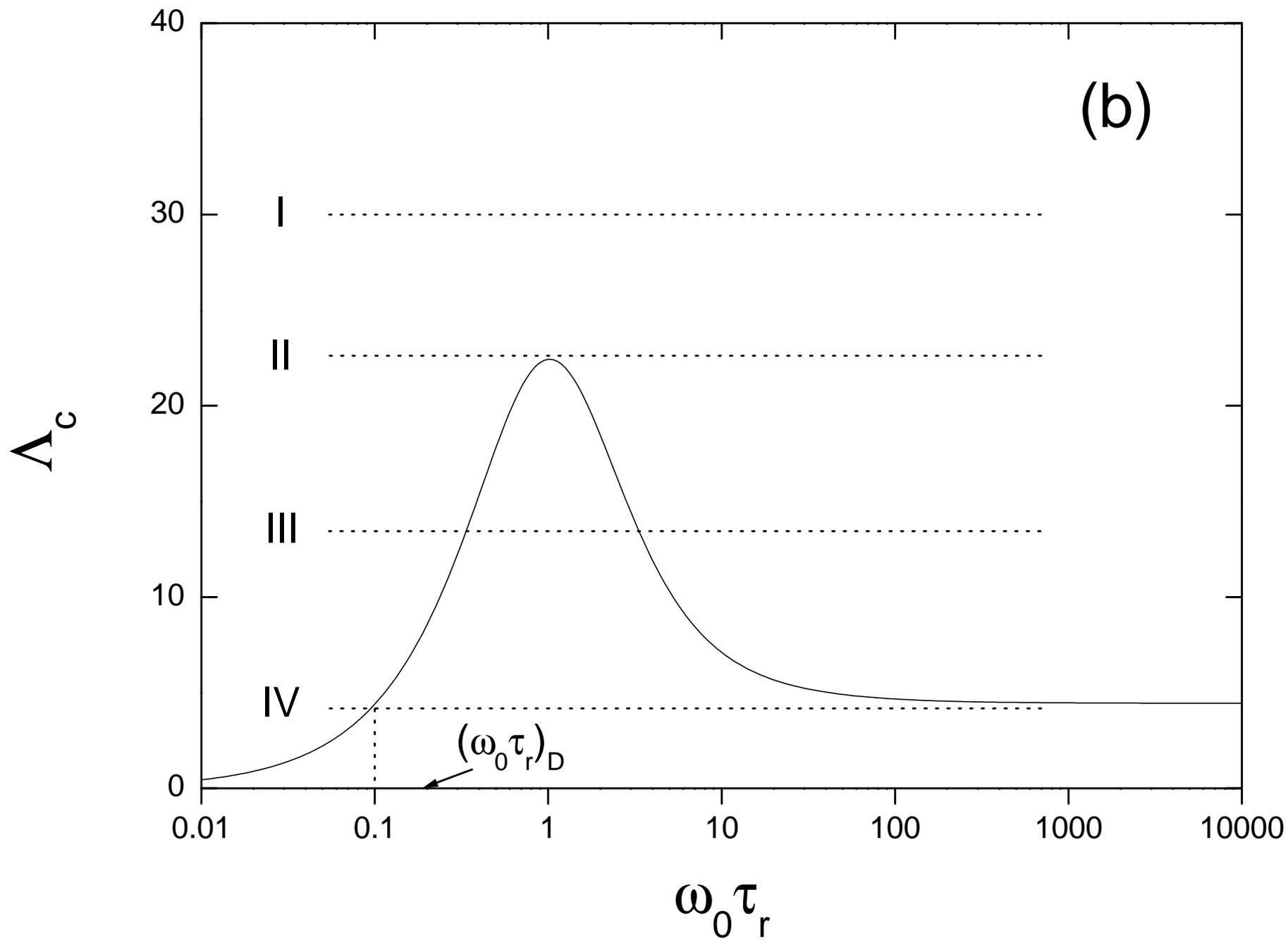
\end{document}